\PassOptionsToPackage{table,dvipsnames}{xcolor}
\documentclass[conference,10pt]{IEEEtran}

\usepackage{amsmath,amssymb}
\usepackage{xifthen}
\usepackage{mathtools}
\usepackage{enumerate}
\usepackage{microtype}
\usepackage{xspace}
\usepackage{bm}
\usepackage[T1]{fontenc}
\usepackage{fancyhdr}
\usepackage{lastpage}
\usepackage{bbm}

\newcommand{\mbs}[1]{\pmb{#1}}
\newcommand{\vect}[1]{{\lowercase{\mbs{#1}}}}
\newcommand{\mat}[1]{{\uppercase{\mbs{#1}}}}

\newcommand{\T}{{\scriptscriptstyle\mathsf{T}}}

\renewcommand{\Re}[1][]{\ifthenelse{\isempty{#1}}{\operatorname{Re}}{\operatorname{Re}\left(#1\right)}}
\renewcommand{\Im}[1][]{\ifthenelse{\isempty{#1}}{\operatorname{Im}}{\operatorname{Im}\left(#1\right)}}

\newcommand{\ev}{\vect{e}}

\newcommand{\ellv}{\vect{\ell}}

\newcommand{\tv}{\vect{t}}

\newcommand{\xv}{\vect{x}}
\newcommand{\yv}{\vect{y}}
\newcommand{\zv}{\vect{z}}

\newcommand{\Fr}{\mathrm{F}}

\newcommand{\Sr}{\mathrm{S}}

\newcommand{\piv}{\vect{\pi}}
\newcommand{\tauv}{\vect{\tau}}

\newcommand{\Lm}{\mat{l}}

\newcommand{\Lc}{{\mathcal L}}

\newcommand{\Nc}{{\mathcal N}}

\newcommand{\EE}{\mathbb{E}}

\newcommand{\RR}{\mathbb{R}}

\newcommand{\Is}{\mat{\mathsf{I}}}
\newcommand{\Ts}{\mat{\mathsf{T}}}

\newcommand{\Ps}{\mat{\mathsf{P}}}
\newcommand{\Qs}{\mat{\mathsf{Q}}}

\newcommand{\CN}[1][]{\ifthenelse{\isempty{#1}}{\mathcal{N}_{\mathbb{C}}}{\mathcal{N}_{\mathbb{C}}\left(#1\right)}}

\renewcommand{\P}[1][]{\ifthenelse{\isempty{#1}}{\mathbb{P}}{\mathbb{P}\left(#1\right)}}
\newcommand{\E}[1][]{\ifthenelse{\isempty{#1}}{\mathbb{E}}{\mathbb{E}\left[#1\right]}}
\newcommand{\I}[1][]{\ifthenelse{\isempty{#1}}{\mathbb{I}}{\mathbb{I}\left\{#1\right\}}}
\renewcommand{\det}[1][]{\ifthenelse{\isempty{#1}}{\mathrm{det}}{\mathrm{det}\left(#1\right)}}
\newcommand{\trace}[1][]{\ifthenelse{\isempty{#1}}{{\rm tr}}{\mathrm{tr}\left(#1\right)}}
\newcommand{\rank}[1][]{\ifthenelse{\isempty{#1}}{\mathrm{rank}}{\mathrm{rank}\left(#1\right)}}
\newcommand{\diag}[1][]{\ifthenelse{\isempty{#1}}{\mathrm{diag}}{\mathrm{diag}\left(#1\right)}}
\newcommand{\Cov}[1][]{\ifthenelse{\isempty{#1}}{\mathsf{Cov}}{\mathsf{Cov}\left(#1\right)}}

\newcommand{\defeq}{\triangleq}

\newtheorem{lemma}{Lemma}
\newtheorem{simplification}{Simplification}

\newcounter{enumi_saved}
\usepackage{answers}
\Newassociation{solution}{Solution}{solutionfile}

\AtBeginDocument{\Opensolutionfile{solutionfile}[\jobname]}
\AtEndDocument{\Closesolutionfile{solutionfile}\clearpage
}

\IfFileExists{MinionPro.sty}{
}{
}

\pdfoutput=1
\usepackage{subcaption}

\usepackage{graphicx,psfrag}
\usepackage{multirow}
\usepackage{array}
\usepackage{setspace}
\usepackage{bbm,mathtools}
\usepackage{amssymb}
\usepackage{color}
\usepackage{enumitem}
\usepackage{stfloats}
\usepackage[flushleft]{threeparttable}
\usepackage{pgfplots}
\usepackage{booktabs}
\usepackage{makecell}
\usepackage{cite}
\PassOptionsToPackage{hyphens}{url}\usepackage[hidelinks]{hyperref}
\usepackage[table,dvipsnames]{xcolor}
\pgfplotsset{minor grid style={dotted}}
\pgfplotsset{major grid style={dashed}}
\usepackage{tikz}
\usetikzlibrary{fit,positioning,arrows,shapes,shapes.multipart,calc,arrows.meta,shapes.geometric,shapes.misc}
\usetikzlibrary{decorations.pathreplacing,angles,quotes,calligraphy}
\usetikzlibrary{automata,fit,positioning,arrows,shapes,shapes.multipart,calc,arrows.meta}
\usepackage[ruled,vlined,linesnumbered]{algorithm2e}
\usepackage[toc,acronym]{glossaries}

\usepackage[left=0.625in,right=0.625in,top=0.7in,bottom=1in]{geometry}
\makeindex

\allowdisplaybreaks

\usepackage{grffile}
\pgfplotsset{compat=newest}
\usetikzlibrary{plotmarks}
\usetikzlibrary{arrows.meta}
\usepgfplotslibrary{patchplots}

\newcommand{\of}[1]{^{(#1)}}

\newcommand{\ind}[1]{{\mathbbm{1}{\{#1\}}}}

\renewcommand{\E}[2][]{\mathbb{E}_{#1}\!\left[#2\right]}
\renewcommand{\P}[2][]{\mathbb{P}_{#1}\!\left[#2\right]}

\renewcommand{\defeq}{=}

\newcommand{\refresh}{{\Sr}}
\newcommand{\fail}{{\Fr}}

\newlength{\rx}
\newlength{\ry}

\DeclareMathOperator*{\argmax}{arg\,max}
\DeclareMathOperator*{\argmin}{arg\,min}

\newcommand{\hoang}[1]{{#1}} 
 
\newcommand{\cgd}[1]{{#1}}  

\makeatletter
\newcommand{\removelatexerror}{\let\@latex@error\@gobble}
\makeatother

\title{Age of Information in Slotted ALOHA \\ With Energy Harvesting\vspace{-.25cm}}

\author{
	\IEEEauthorblockN{Khac-Hoang Ngo\IEEEauthorrefmark{1}, Giuseppe Durisi\IEEEauthorrefmark{1}, Alexandre Graell i Amat\IEEEauthorrefmark{1}, Andrea Munari\IEEEauthorrefmark{2}, and Francisco L\'azaro\IEEEauthorrefmark{2}}
	\IEEEauthorblockA{\IEEEauthorrefmark{1}Department of Electrical Engineering, Chalmers University of Technology, 41296 Gothenburg, Sweden}
	\IEEEauthorblockA{\IEEEauthorrefmark{2}Institute for Communications and Navigation, German Aerospace Center (DLR), 82234 We{\ss}ling, Germany}
	\thanks{This project has received funding from the European Union's Horizon 2020 research and innovation programme under the Marie Skłodowska-Curie grant agreement No 101022113, and from the Swedish Research Council under grant 2021-04970.
	}
	\vspace{-.87cm}
}

\newacronym{AWGN}{AWGN}{additive white Gaussian noise}
\newacronym{MAC}{MAC}{multiple access channel}
\newacronym{UMRA}{UMRA}{unsourced massive random access}
\newacronym{SIMO}{SIMO}{single-input multiple-output}
\newacronym{SISO}{SIMO}{single-input single-output}
\newacronym{iid}{i.i.d.}{independent and identically distributed}
\newacronym{ML}{ML}{maximum likelihood}
\newacronym{PEP}{PEP}{pair-wise error probability}
\newacronym{LLR}{LLR}{log-likelihood ratio}
\newacronym{SNR}{SNR}{signal-to-noise ratio}
\newacronym{SINR}{SINR}{signal-to-interference-plus-noise ratio}

\newacronym{AoI}{AoI}{age of information}
\newacronym{AVP}{AVP}{age-violation probability}
\newacronym{PMF}{PMF}{probability mass function}
\newacronym{CDF}{CDF}{cummulative distribution function}
\newacronym{SA}{SA}{slotted ALOHA}
\newacronym{IRSA}{IRSA}{irregular repetition slotted ALOHA}
\newacronym{SIC}{SIC}{successive interference cancellation}
\newacronym{PLR}{PLR}{packet loss rate}
\newacronym{DE}{DE}{density evolution}
\newacronym{IoT}{IoT}{Internet of Things}
\newacronym{EH}{EH}{energy harvesting}
\newacronym{CP}{CP}{contention period}

\newacronym{wp}{w.p.}{with probability}

\newacronym{BU}{BU}{best-effort uniform}
\newacronym{FC}{FC}{full charge}

\IEEEoverridecommandlockouts
\begin{document}
	
	\maketitle
	\begin{abstract} 
		We examine the age of information (AoI) of a status update system that incorporates energy harvesting and uses the slotted ALOHA protocol.  
		We derive analytically the average AoI and the probability that the AoI exceeds a given threshold. Via numerical results, we investigate two strategies to minimize the \gls{AoI}: transmitting a new update whenever possible to exploit every chance to reduce the \gls{AoI}, and transmitting only when sufficient energy is available to increase the chance of successful delivery. The two strategies are beneficial for low and high update generation rates, respectively. However, an optimized approach that balances the two strategies outperforms them significantly in terms of both AoI and throughput.
	\end{abstract}
	
\section{Introduction} \label{sec:intro}
In delay-sensitive \gls{IoT} applications, devices need to deliver timely status updates to a central gateway. To measure the freshness of status updates, the \gls{AoI} metric has been introduced (see, e.g.,~\cite{Yates2021AoI} and references therein). It captures the time elapsed since the generation of the last update available at the gateway. Recent studies have characterized the AoI for random-access medium sharing protocols, such as slotted ALOHA~\cite{Yates2017,Munari2022_retransmission} and its \cgd{modern} variations~\cite{Munari2020modern,Hoang2021AoI,Munari2023dynamic}. 


IoT devices are designed for low-power, long-term operation and can be placed in remote or hard-to-reach locations, hindering battery replacement. A solution to these challenges is energy harvesting, which allows IoT devices to capture and convert energy from the environment into electrical energy~\cite{Kamalinejad2015wireless}. \hoang{The \gls{AoI} of energy-harvesting devices has been analyzed mainly for the single-source scenario~\cite{Wu2018optimal,Feng2021}.} Existing analyses of ALOHA-based random-access protocols with energy harvesting focused on stability~\cite{Ibrahim2016} and throughput~\cite{Choi2019,Demirhan2019}. Compared to the setting in~\cite{Yates2017,Munari2022_retransmission,Munari2020modern,Hoang2021AoI,Munari2023dynamic}, energy harvesting introduces new factors that significantly affect information freshness, such as the level of available energy at the devices at the time of update generation, and the need for the devices to spend time harvesting energy. However, the impact of energy harvesting on the \gls{AoI} in random-access protocols remains widely unexplored.
 
This paper characterizes the behavior of the AoI in a slotted-ALOHA status update system with energy harvesting. We model energy hervesting as independent Bernoulli processes. 
We assume that each device receives readings from a sensor, and thus cannot generate fresh updates at will. Upon \cgd{receiving} a \hoang{new reading}, the device transmits the update with a probability adapted to its battery level. 
\hoang{A transmitted update is correctly decoded with a probability \cgd{that depends on the transmit power and the level of interference from other devices}.}
 By means of a Markovian analysis, we 
derive the average \gls{AoI} analytically for a given transmission probability. We further provide an approximate analysis 
that results in easy-to-compute and accurate approximations of both the average \gls{AoI} and the \gls{AVP}, \cgd{which is} the probability that the \gls{AoI} exceeds a given threshold.

\hoang{In \cgd{our} numerical experiments, we assume that each slot comprises multiple uses of an \gls{AWGN} channel. We consider both decoding without capture, where the receiver \cgd{performs decoding only in slots containing a single update, and decoding with capture, where the receiver attempts decoding in each slot using \gls{SIC} to recover colliding packets.}}
We investigate the importance of optimally adapting the transmission probability to the available energy. 
On the one hand, transmitting a new update whenever possible \hoang{(\cgd{a.k.a.} best-effort uniform policy~\cite{Wu2018optimal,Feng2021})} exploits every opportunity to reduce the AoI, but increases channel traffic and the risk of losing the update if the transmit power is insufficient. On the other hand, transmitting only with high power increases the chance of successful delivery, but requires devices to ignore some updates while harvesting enough energy. Taking these two strategies as baselines, we compare them with the optimized transmission probability that minimizes the average AoI, minimizes the \gls{AVP}, or maximizes the throughput. \cgd{Numerical results show that significant gains in all three metrics are achieved with the optimized strategy for both decoding with and without capture. Transmitting an update whenever possible is close to optimal for low update generation rates but performs poorly for high update generation rates. As the update generation rate increases, transmitting only when the battery is full has a decreasing gap to the optimal performance without capture. However, this strategy does not benefit from decoding with capture. Furthermore, the strategy optimized for throughput entails a loss in average \gls{AoI} and \gls{AVP}, especially for high update generation rates. Finally, decoding with capture outperforms significantly decoding without capture for the optimized strategy.}

\subsubsection*{Notation}
We \cgd{let} $[m:n] = \{m,m+1,\dots,n\}$, 
$[n] \defeq [1:n]$, and $x^+ = \max\{0,x\}$. 
We denote by $\Is_m$ the $m\times m$ identity matrix, $\mathbf{0}_m$ the $m\times 1$  all-zero vector, $\mathbf{1}_m$ the $m\times 1$ all-one vector, 
$[\xv]_1$ the first entry of $\xv$, and $\ind{\cdot}$ the indicator function.  

\subsubsection*{Reproducible Research} The Matlab code used to
generate our numerical results is available at: \href{https://github.com/khachoang1412/AoI_slottedALOHA_energyHarvesting}{github.com/khachoang1412/AoI\_slottedALOHA\_energyHarvesting}.

\section{System Model} \label{sec:model}
We consider a system with $U$ devices attempting to deliver time-stamped status updates (also called packets) to a \hoang{gateway} through a wireless channel. \cgd{We assume that the updates are generated independently across devices}. Time is slotted and the devices are slot-synchronous. 
Each update transmission spans a slot. A device \hoang{receives a new sensor reading} at the beginning of each slot \gls{wp} $\alpha$. 
	
\subsubsection{Energy Harvesting}
Each device is equipped with a rechargeable battery with capacity $E$ energy units. The devices harvest energy from the environment to recharge their batteries. 
In each slot, one energy unit is harvested by a device \gls{wp}~$\eta$, independently of the other slots and other devices. 
If the battery is full, the device pauses harvesting. 
We denote by $\nu_b$ (computed in Section~\ref{sec:steady_state}) the steady-state probability that the battery level of an arbitrary device is $b \in [0:E]$. 

\subsubsection{Medium Access Protocol} 
The devices access the medium following the slotted ALOHA protocol. 
Specifically, if a device has a new update in a slot, it transmits the update \gls{wp} $\pi_b$ if its battery level is $b$. Obviously, $\pi_0 = 0$, while $\piv = (\pi_1, \dots, \pi_E)$ is a design parameter. \hoang{We assume that the devices always spend all available energy to transmit a packet.}\footnote{\hoang{We shall address the general case where the devices transmit using \cgd{only} part of the available energy in an extension of this paper.}}   
Furthermore, as in~\cite{Ibrahim2016,Choi2019}, we assume that the devices can either transmit or harvest energy in a slot. No feedback is provided by the receiver. 

Consider a device that transmits with $b$ energy units in a slot \cgd{and assume that} the \textit{battery profile} of the remaining devices is $\Lm = (L_0, \dots, L_E)$, i.e., $L_i$ out of the remaining $U-1$ devices have battery level $i \in [0:E]$. 
We denote by $w_{b,\Lm}$ the probability that an \hoang{update transmitted with $b$ energy units is correctly decoded when the battery profile of the other devices is $\Lm$. The dependency of $w_{b,\Lm}$ on $b$ and $\Lm$ captures the impact of the transmit power and \cgd{of} the interference from \cgd{the} other devices. All analytical results in the paper hold for general $w_{b,\Lm}$, while in Section~\ref{sec:results}, we shall instantiate $w_{b,\Lm}$ by considering an \gls{AWGN} channel.} At steady state, the average successful delivery probability of a device that transmits with $b$ energy units is
\begin{equation} \label{eq:avg_w}
    \bar{w}_b = \E[\Lm]{w_{b,\Lm}}
\end{equation}
where $\Lm$ follows the multinomial distribution with number of trials $U-1$, number of events $E+1$, and event probabilities $\{\nu_i\}_{i = 0}^E$. 
The average throughput, i.e., the average number of packets decoded per slot, is given by
	$
    S = \alpha U \sum_{b = 0}^E \nu_b \pi_b \bar{w}_b.$ 

\begin{figure}[t!]
    \centering
    \scalebox{.8}{\begin{tikzpicture}[xscale=0.35,yscale=0.3,domain=0:25,samples=400]
            \draw[-latex] (0,0) -- (22,0) node[below] {$t$};
            \draw[-latex] (0,0) -- (0,9.5) node[left,yshift=-.15cm] {$\delta(t)$};
            
            \draw[thick] (2,3.5) -- (5,6.5) -- (5,1) -- (9,5) -- (9,1) -- (17,9) -- (17,1) -- (20,4);
            \draw[dashed] (0,1) -- (21,1);
            \node at (1,3) () {$\dots$};
            \node at (21,3) () {$\dots$};
            \node at (-.5,1) () {$1$};
            \node at (-.5,7) () {$\theta$};
            \draw[dashed] (0,7) -- (21,7);
    
            \draw [decorate, decoration = {calligraphic brace}] (17,-.2) -- node[below=.1cm,midway] {$Y$} (9,-.2);
    
            \draw [decorate, decoration = {calligraphic brace}] (17,6.7) -- node[below=.1cm,pos=.8] {\footnotesize $Y\!-\!\theta \!+\! 1$} (15,6.7);
        \end{tikzpicture}
    }
	\vspace{-.15cm}
    \caption{Example of the AoI process.}
    \label{fig:AoI_process}
    \vspace{-.5cm}
\end{figure}
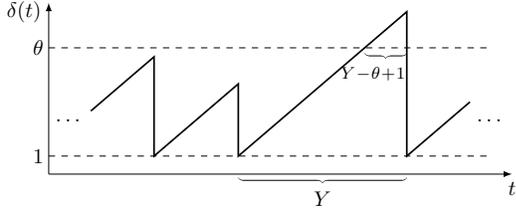

\subsubsection{Age of Information} \label{sec:AoI}
We define the \gls{AoI} of a generic device at slot $t$ as
$\delta(t) \defeq t - \tau(t)$,
where $\tau(t)$ denotes the timestamp of the last received update from this device as of slot $t$. The corresponding stochastic process is denoted as $\Delta(t)$.  
The \gls{AoI} grows linearly with time and is reset to $1$ when a new update is successfully decoded. It has a saw-tooth shape as illustrated in Fig.~\ref{fig:AoI_process}. We are interested in the average \gls{AoI}
$
    \bar{\Delta} = \E{\Delta(t)} 
$
and the \gls{AVP}
$
    \zeta(\theta) = \P{\Delta(t) > \theta}.
$

\section{\hoang{Battery Evolution}} \label{sec:steady_state}



\subsubsection{Battery Level of a Generic Device}
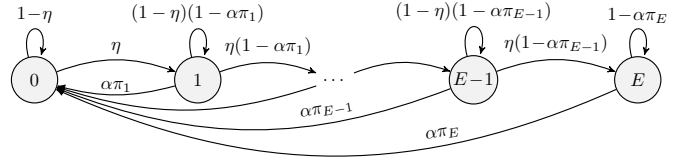
\begin{figure}[t!]
    \centering
    \vspace{-.3cm}
    \begin{tikzpicture}
        \tikzset{node distance=3cm, 
         every state/.style={ 
           semithick,
           fill=gray!10},
         initial text={},     
         double distance=4pt, 
         every edge/.style={  
         draw,
           ->,>=stealth',     
           auto,
           semithick}}
        \hspace{-.1cm}
        \scalebox{.7}{
        \node[state] (b0) {$0$};
        \node[state, right=2.2cm of b0] (b1) {$1$};
        \node[right=1.8cm of b1] (mid) {$\dots$};
        \node[state, right=1.8cm of mid] (b3) {$\!\!E\!-\!1\!\!$};
        \node[state, right=2.2cm of b3] (b4) {$E$};
        
        \draw (b0) edge[loop above] node[above,midway] {$1\!-\!\eta$} (b0);
        \draw (b1) edge[loop above] node[above,midway] {$(1-\eta)(1-\alpha \pi_1)$} (b1);
        \draw (b3) edge[loop above] node[above,midway] {$(1-\eta)(1-\alpha \pi_{E-1})$} (b3);
        
        \draw (b0) edge[bend left=20] node[above,midway] {$\eta$} (b1);
        \draw (b1) edge[bend left=20] node[above,midway] {$\eta(1-\alpha \pi_1)$} (mid);
        \draw (mid) edge[bend left=20] node[above,midway] {} (b3);
        \draw (b3) edge[bend left=20] node[above,pos=.5] {$\eta(1\!-\!\alpha \pi_{E-1})$} (b4);
        \draw (b4) edge[loop above] node[above,midway] {$1\!-\!\alpha \pi_{E}$} (b4);

        \draw (b1) edge[bend left=15] node[above=-.07cm,midway] {$\alpha \pi_1$} (b0);
        \draw (mid) edge[bend left=18] node[above,midway] {} (b0);
        \draw (b3) edge[bend left=21] node[above,pos=.3,rotate=10] {$\alpha \pi_{E-1}$} (b0);
        \draw (b4) edge[bend left=24] node[above,pos=.3,rotate=10] {$\alpha \pi_{E}$} (b0);
        }
    \end{tikzpicture}
    \vspace{-1.2cm}
    \caption{Markov chain $M_1$ describing the battery level of a device.}
    \label{fig:markov_battery}
    \vspace{-.3cm}
\end{figure}

The evolution of the battery level of a generic device is captured by the Markov chain $M_1$ shown in Fig.~\ref{fig:markov_battery}. 
Each state represents a battery level. The transition probabilities between the states can be readily computed. Specifically, a device in state $0$ cannot transmit, thus it either remains in this state if it does not harvest energy (\gls{wp} $1-\eta$) or jumps to state $1$ if an energy unit arrives (\gls{wp} $\eta$). A device in state $i \in [E]$ moves to state $0$ if it generates and transmits a new update (\gls{wp} $\alpha \pi_i$). Otherwise, if $i<E$, the device either remains in state $i$ if no energy is harvested (\gls{wp} $(1-\eta)(1-\alpha \pi_i)$) or jumps to state $i+1$ if an energy unit \cgd{is harvested}  (\gls{wp} $\eta(1-\alpha \pi_i)$). If the battery is full, i.e., $i = E$, the device remains in state $E$ if it does not transmit (\gls{wp} $1-\alpha \pi_i$). 
From these transition probabilities, we compute the steady-state distribution $\{\nu_b\}_{b = 0}^E$ 
by solving the balance equations.

\subsubsection{Battery Profile of $U-1$ Devices}
The battery profile $\Lm$ of \cgd{the other} $U-1$ devices can take values in
$
    \Lc = \big\{(\ell_0, \ell_1,\dots, \ell_E) \colon \sum_{i = 0}^E \ell_i = U\!-\!1, \ell_i \in [0:U\!-\!1], i \in [0:E]\big\}.
$
We now describe the evolution of $\Lm$ within a slot. Let $\ellv' = (\ell'_0, \ell'_1,\dots, \ell'_E)$ and $\ellv = (\ell_0,\ell_1,\dots, \ell_E)$ denote the states at the beginning and the end of a slot, respectively. Let also $u_{j,k}$ be the number of devices whose battery goes from level $j$ to level $k$. We have that 
 \begin{align}
     u_{j,k} &\in [0:\min\{\ell'_j,\ell_k\}], \quad j,k \in [0:E], \label{eq:tmp283} \\
     \ell'_0 &= u_{0,0} + u_{0,1}, \\
     \ell'_i &= u_{i,i} + u_{i,0} + u_{i,i+1}, \quad i\in [1:E-1], \\
     \ell'_E &= u_{E,E} + u_{E,0}, \\
     \ell_0 &= u_{0,0} + \textstyle\sum_{i=1}^E u_{i,0},  \\
     \ell_i &= u_{i,i} + u_{i-1,i}, \quad i\in [E]. \label{eq:tmp288}
 \end{align}
\cgd{hence,} the transition probability \cgd{$\P{\ellv' \to \ellv}$} is 
\begin{align} \label{eq:trans_profile}
    &\P{\ellv' \to \ellv} = \sum_{\{u_{j,k}\} \colon \text{\eqref{eq:tmp283}--\eqref{eq:tmp288}}} \bigg(\prod_{j,k \in [0:E]} p_{j,k}^{u_{j,k}}\bigg) \notag \\
    &\qquad \cdot \binom{\ell'_0}{u_{0,0}} \binom{\ell'_E}{u_{E,0}}\prod_{j=1}^{E-1} \binom{\ell'_j}{u_{j,0}} \binom{\ell'_j - u_{j,0}}{u_{j,j}},
\end{align}
where $p_{j,k}$ is the transition probability from state $j$ to state $k$ of the Markov chain $M_1$ in Fig.~\ref{fig:markov_battery}.

\section{AoI Analysis} \label{sec:AoI}
We now derive the average \gls{AoI} of a generic device. 

\subsubsection{Preliminaries}
We denote by $B\of{s}$ the battery level of the device of interest at the end of slot $s$. We let $X\of{s} = \hoang{\refresh}$ \cgd{(standing for ``success'')} if the device successfully delivers an update in the slot, and \cgd{$X\of{s} = \fail$ (standing for ``fail'')} otherwise. 
Furthermore, we denote the battery profile of the remaining $U\!-\!1$ devices at the end of slot $s$ 
by $\Lm\of{s} = (L\of{s}_0, \dots, L\of{s}_E)$. Consider an ancillary Markov chain 
\cgd{$Z\of{s} = (X\of{s},B\of{s},\Lm\of{s})$}. We next derive the transition probability from state \cgd{$(x', b',\ellv')$} to state \cgd{$(x, b,\ellv)$}. If $X\of{s-1}=\refresh$, the device of interest depletes its battery and thus cannot transmit an update in slot~$s$. Therefore, 
\begin{align}
    &\P{(\refresh, 0,\ellv') \to (x,b,\ellv)} = \ind{x = \refresh}   \notag \\ 
    &\quad \cdot((1-\eta) \ind{b \!=\! 0} + \eta \ind{b \!=\! 1} ) \P{\ellv' \to \ellv},
\end{align}
where $\P{\ellv' \to \ellv}$ is given in \eqref{eq:trans_profile}.
If $X\of{s-1} \!=\! \fail$, we separate the cases $X\of{s} \!=\! \refresh$ and $X\of{s} \!=\! \fail$. First, $X\of{s} \!=\! \refresh$ if in slot $s$ the device generates and transmits a new update (\gls{wp} $\alpha \pi_{b'}$), and the update is successfully decoded (\gls{wp} $w_{b',\ellv'}$). 
In this case, the device depletes its battery after slot $s$. Therefore, 
\begin{equation}
    \!\P{(\fail, {b'},\ellv') \!\to\! (\refresh, b,\ellv)} =  \alpha \pi_{b'} w_{b',\ellv'} \ind{b \!=\! 0} 
    \P{\ellv' \!\to\! \ellv}\!.
\end{equation}
Second, $X\of{s} \!=\! \fail$ if in slot $s$ the device either does not transmit or transmits but fails to deliver the packet. It follows that
\begin{align}
    &\P{(\refresh, {b'},\ellv')\rightarrow (\fail, b,\ellv)} \notag \\ 
    &= \big[(1 - \alpha \pi_{b'} \ind{b' > 0}) \big((1-\eta)\ind{b' = b < E} \notag \\
    &\qquad\quad + \eta \ind{b = b' + 1} + \ind{b=b'=E}\big) \notag \\
    &\qquad + \alpha \pi_{b'} (1-w_{b',\ellv'}) \ind{b = 0}\big]  \P{\ellv' \to \ellv}.
\end{align}

\subsubsection{Average AoI}
As shown in Fig.~\ref{fig:AoI_process}, we denote by $Y$ the {\em inter-refresh} time, i.e., the number of slots that elapse between two successive status updates for the device of interest. Right after a refresh, the current \gls{AoI} is set to $1$. By proceeding as in \cite[Sec.~II-A]{Yates2021AoI} or~\cite[Sec.~III]{Munari2022_retransmission}, we observe that the average \gls{AoI} can be expressed in terms of the moments of $Y$ as 
\begin{equation} \label{eq:avgAoI_via_momentsY}
    \bar{\Delta} = 1 + \frac{\E{Y^2}}{2\E{Y}}.
\end{equation}
We next derive the moments of $Y$. 
Without loss of generality, we assign index $1$ to the first slot contributing to the current inter-refresh time. We expand $\E{Y}$ as
\begin{align}
    \E{Y} &= \sum_{x \in \{\fail,\refresh\}} \sum_{b \in [0:E]} \sum_{\ellv \in \Lc} \E{Y\vert Z\of{1} = (x,b,\ellv)} \notag \\
    &\qquad \cdot \mathbb{P}\big[Z\of{1} = (x,b,\ellv)\big]. \label{eq:EY_exact}
\end{align}
To compute $\P{Z\of{1} = (x,b,\ellv)}$, we note that the state \cgd{at the end of} a slot with \gls{AoI} refresh is of the form $(\refresh, 0,\ellv)$, and \cgd{the state at the end of} slot~$1$ can only be $(\fail, 0,\ellv)$ or $(\fail, 1,\ellv)$. Therefore,
\begin{align} 
    &\mathbb{P}\big[Z\of{1} = (x,b,\ellv)\big] = \ind{x = \fail, b \in \{0,1\}} \notag \\
    &\qquad \cdot \frac{\sum_{\ellv' \in \Lc}\P{(\refresh, 0,\ellv') \to (\fail, b,\ellv)}}{\sum_{b \in \{0,1\}, \ellv \in \Lc}\sum_{\ellv' \in \Lc}\P{(\refresh, 0,\ellv') \to (\fail, b,\ellv)}}. \label{eq:pZ1}
\end{align}
The conditional expectation $\E{Y\vert Z\of{1} = (x,b,\ellv)}$ can be derived via a first-step analysis~\cite[Sec.~III-4]{TaylKarl98}. If the packet from the device of interest is decoded in slot $1$, i.e., $X\of{s} = 1$, 
the inter-refresh time is $1$. It follows that 
\begin{equation}
    \mathbb{E}\big[Y | Z\of{1} = (\refresh, b,\ellv)\big] = 1, \quad b \in [0:E], \ellv \in \Lc. \label{eq:tmp385}
\end{equation}
If $Z\of{1} = (\fail, b,\ellv)$, the inter-refresh time can be computed as the sum of the number of slots until a transmitted packet is successfully decoded. This can be conveniently computed by conditioning on the outcome of the first transition. Specifically, 
we define $r(b,\ellv) = \sum_{b''\in [0:E], \ellv'' \in \Lc} \P{(\fail, b,\ellv) \to (\refresh, {b''},\ellv'')}$, $q((b,\ellv) \!\to\! (b'',\ellv'')) \!=\! \P{(\fail, b,\ellv) \!\to\! (\fail, {b''},\ellv'')}$, and proceed as
\begin{align}
    &\EE\big[Y | Z\of{1} = (\fail, b,\ellv)\big] = \notag \\
    &\quad 1 + \sum_{z\in \{\fail,\refresh\} \times [0:E] \times \Lc} \E{Y | Z\of{1} = z} \P{(\fail, b,\ellv) \to z} \label{eq:tmp390}\\
    &= 1 + r(b,\ellv) \notag \\
    &\quad + \sum_{b''\in [0:E], \ellv'' \in \Lc} \!\!\E{Y | Z\of{1} \!=\! (\fail, {b''},\ellv'')} \notag \\
    &\qquad\qquad \qquad \cdot q((b,\ellv) \to (b'',\ellv'')). \label{eq:tmp393}
\end{align}
In~\eqref{eq:tmp390}, the Markov property ensures that the average duration, once the transition to state $z$ has occurred, is equal to the one that we would have by starting from such state. Let $\ev$ and $\pmb{r}$ be vectors that contain $\E{Y | Z\of{1} = (\fail, b,\ellv)}$ and $r(b,\ellv)$, respectively, \cgd{for all values of} $(b,\ellv)$. Let $\Qs$ be a matrix that contains $q((b,\ellv) \to (b'',\ellv''))$ for all $(b,\ellv)$ and $(b'',\ellv'')$. The full-rank system of equations obtained from~\eqref{eq:tmp393} can be expressed compactly as 
$(\Is - \Qs) \ev = \mathbf{1} + \pmb{r}$. Therefore, 
    $\ev = (\Is - \Qs)^{-1} (\mathbf{1} + \pmb{r})$. 
Substituting this,~\eqref{eq:pZ1}, and~\eqref{eq:tmp385} into~\eqref{eq:EY_exact}, we obtain $\E{Y}$. 

The second-order moment $\E{Y^2}$ \cgd{can also be} computed via a first-step analysis. This yields
\begin{align}
    &\E{Y^2 | Z\of{1} = (\refresh, b,\ellv)} = 1, \quad b \!\in\! [0:E], \ellv \!\in\! \Lc, \\ \label{eq:tmp404}
    &\E{Y^2 | Z\of{1} = (\fail, b,\ellv)} \notag \\
    &= 1 + 2 \sum_{z\in \{\fail,\refresh\} \times [0:E] \times \Lc} \E{Y | Z\of{1} = z} \P{(\fail, b,\ellv) \to z} \notag\\
    &\quad + \sum_{z\in \{\fail,\refresh\} \times [0:E] \times \Lc} \!\!\! \E{Y^2 | Z\of{1} \!=\! z} \P{(\fail, b,\ellv) \to z} \\
    &= -1 + 2 \E{Y | Z\of{1} = (\fail_b,\ellv)} + r(b,\ellv) \notag \\
    &\quad + \sum_{b''\in [0:E], \ellv'' \in \Lc} \E{Y^2 | Z\of{1} = (\fail, b'',\ellv'')} \notag \\
    &\qquad\qquad \qquad \cdot q((b,\ellv) \!\to\! (b'',\ellv'')). \label{eq:tmp413}
\end{align}
Let $\ev_2$ be a vector that contains $\E{Y^2 | Z\of{1} = (\fail, b,\ellv)}$ \cgd{for all values of} $(b,\ellv)$. We \cgd{can} express~\eqref{eq:tmp413} compactly as $(\Is - \Qs)\ev_2 = -\mathbf{1} + 2\ev+\pmb{r}$. It follows that 
    $\ev_2 = (\Is - \Qs)^{-1} (-\mathbf{1} + 2\ev+\pmb{r})$. 
Using this,~\eqref{eq:pZ1}, and~\eqref{eq:tmp404}, we compute $\E{Y^2}$ via an expansion analogous to~\eqref{eq:EY_exact}. 
Finally, \cgd{we obtain} the average \gls{AoI} $\bar{\Delta}$ by inserting the computed moments of $Y$ into~\eqref{eq:avgAoI_via_momentsY}. 

The above exact computation becomes \cgd{infeasible when} $U$ and $E$ \cgd{are large}. Specifically,  computing the transition probabilities between all $2(E+1)\binom{U+E-1}{E}$ states of the chain $Z$ is prohibitive for large $U$ and $E$. This motivates us to propose an approximate analysis 
in the next section.

\section{Approximate AoI Analysis} \label{sec:AoI_approx}
To simplify the analysis, we ignore the time dependency of the battery profile of the devices whose performance is not tracked. Specifically, we assume the following.
\begin{simplification} \label{simplification}
    given a device of interest, the battery profile $\Lm$ of the remaining $U-1$ devices 
    is \textit{independent} across slots. 
\end{simplification}

This \cgd{simplification} allows us to analyze the behavior of the system, and, as we shall see, results in tight approximations on the average AoI and AVP for all scenarios explored. Under \cgd{this simplification}, the successful delivery probability of a device that transmits with $b$ energy units is the average of $w_{b,\Lm}$ over $\Lm$, \cgd{i.e.,} $\bar{w}_b$  in~\eqref{eq:avg_w}.
\cgd{This allows us to derive} the distribution of the inter-refresh time $Y$ in closed form. 

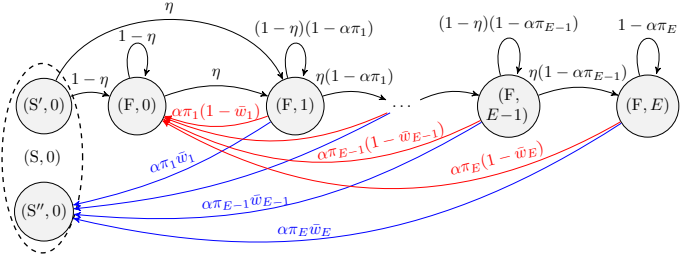
\begin{figure}[t!]
    \centering
    \vspace{-.8cm}
    \begin{tikzpicture}
        \tikzset{node distance=2.6cm, 
         every state/.style={ 
           semithick,
           fill=gray!10},
         initial text={},     
         double distance=4pt, 
         every edge/.style={  
         draw,
           ->,>=stealth',     
           auto,
           semithick}}
       \hspace{-1.1cm}
        \scalebox{.63}{
        \node[state] at (0,0) (b0) {$(\fail,0)$};
        \node[state, left=.75cm of b0] (R) {$\!(\refresh',0)\!$};
        \node[state, right=2.1cm of b0] (b1) {$(\fail,1)$};
        \node[right=1.3cm of b1] (mid) {$\dots$};
        \node[state, right=1.2cm of mid,align=center] (b3) {$(\fail,$ \\ $\!\!\!{E\!-\!1})\!\!$};
        \node[state, right=1.6cm of b3] (b4) {$(\fail,E)$};
        \node[state, below=1cm of R] (R2) {$\!(\refresh'', 0)\!$};
        
        \node[ellipse,draw = black,dashed, thick,
        minimum width = 1.7cm, 
        minimum height = 4cm] (e) at (-2,-1.1cm) {$(\refresh,0)$};

        \draw (R) edge[bend left=20] node[above=-.01cm,midway] {$1-\eta$} (b0);
        \draw (R) edge[bend left=65] node[above,midway] {$\eta$} (b1);
        \draw[color=blue] (b1) edge[bend left=10] node[above,midway,rotate=18] {$\alpha\pi_1\bar{w}_1$} (R2);
        \draw[color=blue] (mid) edge[bend left=11] node[above,midway] {} (R2);
        \draw[color=blue] (b3) edge[bend left=18] node[above=-.06cm,rotate=5,pos=.6] {$\alpha\pi_{E-1}\bar{w}_{E-1}$} (R2);
        \draw[color=blue] (b4) edge[bend left=23] node[above=-.05cm,pos=.6] {$\alpha\pi_{E}\bar{w}_{E}$} (R2);
        
        \draw (b0) edge[loop above] node[above=-.15cm,midway] {$1-\eta$} (b0);
        \draw (b1) edge[loop above] node[above,xshift=.4cm,midway] {$(1-\eta)(1-\alpha \pi_1)$} (b1);
        \draw (b3) edge[loop above] node[above,midway] {$(1-\eta)(1-\alpha \pi_{E-1})$} (b3);
        
        \draw (b0) edge[bend left=20] node[above,midway] {$\eta$} (b1);
        \draw (b1) edge[bend left=20] node[above,midway] {$\eta(1-\alpha \pi_1)$} (mid);
        \draw (mid) edge[bend left=20] node[above,midway] {} (b3);
        \draw (b3) edge[bend left=20] node[above,midway] {$\eta(1-\alpha \pi_{E-1})$} (b4);
        \draw (b4) edge[loop above] node[above,midway] {$1-\alpha \pi_{E}$} (b4);

        \draw[color=red] (b1) edge[bend left=20] node[above,midway] {$\alpha \pi_1(1-\bar{w}_1)$} (b0);
        \draw[color=red] (mid) edge[bend left=23] node[above,midway] {} (b0);
        \draw[color=red] (b3) edge[bend left=27] node[rotate=12,above=-.01cm,pos=.3] {$\alpha \pi_{E-1}(1-\bar{w}_{E-1})$} (b0);
        \draw[color=red] (b4) edge[bend left=30] node[rotate=14,above,pos=.26] {$\alpha \pi_{E}(1-\bar{w}_{E})$} (b0);
        }
    \end{tikzpicture}
    \vspace{-1.8cm}
    \caption{Markov chain $M_2$ to track the AoI refresh of a device. 
    }
    \label{fig:markov_refresh_absorp}
    \vspace{-.5cm}
\end{figure}

\subsubsection{Distribution of the Inter-Refresh Time $Y$}
To track the behavior of the device of interest, we consider the Markov chain $M_2$ in Fig.~\ref{fig:markov_refresh_absorp}, which is obtained from $M_1$ in Fig.~\ref{fig:markov_battery} as follows. We  split the battery state $0$ into two states: \gls{AoI} refresh $(\refresh, 0)$ and no \gls{AoI} refresh $(\fail,0)$. The state $b \in [E]$ in $M_1$ is called $(\fail,b)$ in $M_2$. The device visits this state if its AoI value is not refreshed and its battery level is $b$. After some manipulations, we obtain the transition probabilities between these states under Simplification~\ref{simplification} as depicted in Fig.~\ref{fig:markov_refresh_absorp}. \hoang{\cgd{We can also interpret} the chain $M_2$ as obtained by grouping the states $\{(x,b,\ellv) \colon \ellv \in \Lc\}$ of $Z\of{s}$ into a single state $(x,b)$, and computing the transition probabilities as $\P{(x',{b'}) \to (x,b)} = \E{\sum_{\ellv \in \Lc} \P{(x',b',\Lm') \to (x,b,\ellv)}}$ where the expectation is over the steady state distribution of $\Lm'$.}
Next, we further split state $(\refresh, 0)$ into two states: $(\refresh, 0')$ (with only outgoing transitions from $(\refresh, 0)$) and $(\refresh'',0)$ (with only incoming transitions to $(\refresh,0)$). 
The chain $M_2$ is a \emph{terminating Markov chain}  with one absorbing state $(\refresh'',0)$ and $E + 1$ transient (i.e., non-absorbing) states $\{(\refresh',0), (\fail,0), (\fail,1),\dots,(\fail,E)\}$. 
Observe that the inter-refresh time $Y$ is the absorption time into $(\refresh'',0)$ when starting from $(\refresh',0)$. The distribution of the time until absorption of a terminating Markov chain is called the {\em discrete phase-type distribution} and 
 has been analyzed in~\cite[Sec.~2.2]{Neuts1994}. Leaning on this result, we characterize the distribution of $Y$ in the next lemma, whose proof is omitted due to the space limitations.

\begin{lemma}[Distribution of the inter-refresh time] \label{lemma:dist_Y}
    Under Assumption~\ref{simplification}, it holds that 
    \begin{align} \label{eq:dist_Y}
        \P{Y = y} &= [\Ts^{y-1}\tv_0]_1, \quad y = 1,2,\dots, \\
        \P{Y \ge y} &= [\Ts^{y-1} \mathbf{1}_{E+1}]_1, \quad y = 1,2,\dots, \label{eq:CDF_Y}
    \end{align}
    where 
        $\tv_0 = [0 ~~ \alpha \pi_1 \bar{w}_1 ~~ \dots ~~ \alpha \pi_E \bar{w}_E]^\T$
    and $\Ts$ is given in~\eqref{eq:T}.
    \begin{figure*}[b!]
	\vspace{-.5cm}
     \begin{align} \label{eq:T} \small
        \Ts = \begin{bmatrix}
            1-\eta & \eta & 0 &  0 & \dots & 0 & 0 \\
            \alpha \pi_1(1-\bar{w}_1) & (1\!-\!\eta) (1\!-\!\alpha \pi_1) & \eta(1-\alpha \pi_1) & 0 & \dots &  0 & 0 \\
            \alpha \pi_2(1-\bar{w}_2) & 0 & (1-\eta) (1-\alpha \pi_2) & \eta(1\!-\!\alpha \pi_2) & \dots &  0 & 0 \\
            \vdots & \vdots & \vdots & \ddots & \ddots & \vdots & \vdots \\
            \alpha \pi_{E-1}(1-\bar{w}_{E-1}) & 0 & 0 & 0 & \dots & (1\!-\!\eta) (1\!-\!\alpha \pi_{E-1}) & \eta(1-\alpha \pi_{E-1}) \\
            \alpha \pi_{E}(1-\bar{w}_{E}) & 0 & 0 & 0 & \dots & 0 & 1-\alpha \pi_{E}
        \end{bmatrix}
    \end{align}
\end{figure*}
    Furthermore, 
    \begin{align} 
        \E{Y} &= [(\Is_{E+1}-\Ts)^{-1} \mathbf{1}_{E+1}]_1, \label{eq:mean_Y} \\
        \E{Y^2} &= 2[(\Is_{E+1}-\Ts)^{-2} \mathbf{1}_{E+1}]_1 - \E{Y}. \label{eq:mean2_Y} 
    \end{align}
\end{lemma}

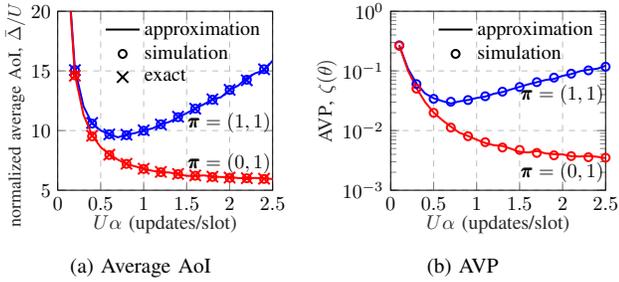
\begin{figure}[t!]
	\centering
    \vspace{-.1cm}
	    
\subcaptionbox{Average AoI 
}{
	\begin{tikzpicture}[scale=.75]
		\begin{axis}[%
			width=1.5in,
			height=1.25in,
			clip mode=individual,
			scale only axis,
			unbounded coords=jump,
			xmin=0,
			xmax=2.5,
			xtick= {0, 0.5, 1, 1.5,2,2.5},
			xlabel style={font=\color{white!15!black},yshift=4pt},
			xlabel={$U \alpha$ (updates/slot)},
			ymin=5,
ymax=20,
ytick={5,10,15,20},
			yminorticks=true,
			ylabel style={font=\color{white!15!black},yshift=-4pt,xshift=-7pt},
			ylabel={\small normalized average AoI, $\bar{\Delta}/U$},
			axis background/.style={fill=white},
			title style={font=\bfseries},
			xmajorgrids,
			ymajorgrids,
			legend style={at={(.99,0.99)}, anchor=north east, legend cell align=left, align=left, fill=none,draw=none}
			]
			
			\addplot [line width = 1,color=blue,forget plot]
			table [x index = {0}, y index={1}, col sep=comma]
			{./fig/sim_vs_analytic_capture_pi11.csv}; 
		
		\addplot [line width = 1,color=blue,mark = o, mark color = blue,only marks,forget plot,mark repeat=2]
		table [x index = {0}, y index={2}, col sep=comma]
		{./fig/sim_vs_analytic_capture_pi11.csv}; 
	
	\addplot [line width = 1,color=blue,mark = x, mark color = blue,only marks,forget plot, mark size = 4,mark repeat=2]
	table [x index = {0}, y index={3}, col sep=comma]
	{./fig/sim_vs_analytic_capture_pi11.csv}; 

\addplot [line width = 1,color=red,forget plot]
table [x index = {0}, y index={1}, col sep=comma,mark repeat=2]
{./fig/sim_vs_analytic_capture_pi01.csv}; 

\addplot [line width = 1,color=red,mark = o, mark color = red,only marks,forget plot,mark repeat=2]
table [x index = {0}, y index={2}, col sep=comma]
{./fig/sim_vs_analytic_capture_pi01.csv}; 

\addplot [line width = 1,color=red,mark = x, mark color = red,only marks,forget plot,mark size = 4,mark repeat=2]
table [x index = {0}, y index={3}, col sep=comma]
{./fig/sim_vs_analytic_capture_pi01.csv}; 

\node at (axis cs:2,10.7) () {$\piv = (1, 1)$};
\node at (axis cs:2,7.3) () {$\piv = (0, 1)$};
    \addplot [line width = 1,color = black]
			table[row sep=crcr]{%
                -1 0 \\ 
                -1 0 \\ 
                };
    \addlegendentry{approximation};
    \addplot [line width = 1,mark = o, mark color = black,only marks]
			table[row sep=crcr]{%
                -1 0 \\ 
                -1 0 \\ 
                };
    \addlegendentry{simulation};
     \addplot [line width = 1,mark = x, mark color = black,only marks,mark size = 4]
			 table[row sep=crcr]{%
                 -1 0 \\ 
                 -1 0 \\ 
                 };
     \addlegendentry{exact};
   \end{axis}
	\end{tikzpicture}}
\subcaptionbox{\gls{AVP} 
}{
	\begin{tikzpicture}[scale=.75]
		\begin{axis}[%
			width=1.5in,
			height=1.25in,
			scale only axis,
			unbounded coords=jump,
			xmin=0,
			xmax=2.5,
			xtick= {0, 0.5, 1, 1.5,2,2.5},
			xlabel style={font=\color{white!15!black},yshift=4pt},
			xlabel={$U \alpha$ (updates/slot)},
			ymode=log,
			ymin=1e-3,
			ymax=1,
			yminorticks=true,
			ylabel style={font=\color{white!15!black},yshift=-4pt},
			ylabel={\gls{AVP}, $\zeta(\theta)$},
			axis background/.style={fill=white},
			title style={font=\bfseries},
			xmajorgrids,
			ymajorgrids,
			legend style={at={(.99,0.99)}, anchor=north east, legend cell align=left, align=left, fill=none,draw=none}
			]
			
			\addplot [line width = 1,color=blue,forget plot]
	table [x index = {0}, y index={4}, col sep=comma]
{./fig/sim_vs_analytic_capture_pi11.csv}; 

\addplot [line width = 1,color=blue,mark = o, mark color = blue,only marks,forget plot,mark repeat=2]
	table [x index = {0}, y index={5}, col sep=comma]
{./fig/sim_vs_analytic_capture_pi11.csv}; 

\addplot [line width = 1,color=red,forget plot]
	table [x index = {0}, y index={4}, col sep=comma]
{./fig/sim_vs_analytic_capture_pi01.csv}; 

\addplot [line width = 1,color=red,mark = o, mark color = red,only marks,forget plot,mark repeat=2]
	table [x index = {0}, y index={5}, col sep=comma]
{./fig/sim_vs_analytic_capture_pi01.csv}; 

\node at (axis cs:2,4e-2) () {$\piv = (1, 1)$};
\node at (axis cs:2,2e-3) () {$\piv = (0, 1)$};
    \addplot [line width = 1,color = black]
table[row sep=crcr]{%
	-1 1 \\ 
	-1 1 \\ 
};
\addlegendentry{approximation};
\addplot [line width = 1,mark = o, mark color = black,only marks]
table[row sep=crcr]{%
	-1 1 \\ 
	-1 1 \\ 
};
\addlegendentry{simulation};
   \end{axis}
\end{tikzpicture}}
	\vspace{-.1cm}
	\caption{Average AoI and \gls{AVP} vs. average total number of new updates  in a slot ($U\alpha$). Here, $U = 30$, $\eta = 0.05$, $E = 2$, $n = 100$, $R = 0.8$, $\theta = 1000$, $\sigma^2 = -20$~dB, and the decoder is with capture. 
	}
	\label{fig:sim_vs_analytic}
	\vspace{-.5cm}
\end{figure}

\subsubsection{Approximate Average \gls{AoI}}
By inserting the moments of $Y$ given in \eqref{eq:mean_Y} and \eqref{eq:mean2_Y} 
into \eqref{eq:avgAoI_via_momentsY}, we obtain the average \gls{AoI} under Simplification~\ref{simplification} as
    \begin{equation} \label{eq:avgAoI}
        \bar{\Delta} = \frac{1}{2} + \frac{[(\Is_{E+1}-\Ts)^{-2} \mathbf{1}_{E+1}]_1}{[(\Is_{E+1}-\Ts)^{-1} \mathbf{1}_{E+1}]_1}.
    \end{equation}

\subsubsection{Approximate \gls{AVP}}
Without loss of generality, we start tracking the process (i.e., we set $t \!=\! 0$) right after the first \gls{AoI} refresh, which is indexed as the $0$th refresh. Let $t_i$ be the time instant of the $i$th \gls{AoI} refresh and $y_i = t_i - t_{i-1}$ the duration of the $i$th inter-refresh period. Using~\cite[Eq.~(15)]{Yates2021AoI}, we compute the \gls{AVP} under Simplification~\ref{simplification} as
\begin{align}
    \zeta(\theta) &= \lim_{T \to \infty} \frac{1}{T}\int_{0}^T \ind{\Delta(t) > \theta} {\rm d} t \\
    &= \lim_{m \to \infty} \frac{1}{\sum_{i=1}^m y_i} \sum_{i=1}^m \int_{t_{i-1}}^{t_i} \ind{\Delta(t) > \theta} {\rm d} t \\
    &= \lim_{m \to \infty} \frac{1}{\sum_{i=1}^m y_i} \sum_{i=1}^m (y_i - \theta + 1)^+ \label{eq:tmp648} \\
    &= \lim_{m \to \infty} \frac{1}{\frac{1}{m}\sum_{i=1}^m y_i} \sum_{y=0}^\infty \frac{|\{i \in [m]\colon y_i = y\}|}{m}(y \!-\! \theta \!+\! 1)^+ \notag \\
    &= \frac{1}{\E{Y}} \sum_{y=0}^\infty \P{Y=y}(y - \theta + 1)^+ \label{eq:tmp649}\\
    &= \frac{1}{\E{Y}} \bigg(\sum_{y=\theta}^\infty y \P{Y=y} - (\theta\!-\!1) \sum_{y = \theta}^\infty \P{Y=y} \bigg) \\
    &= 1 - \frac{1}{\E{Y}} \bigg( \sum_{y=1}^{\theta-1} y \P{Y=y} - (\theta\!-\!1) \P{Y\ge \theta} \bigg) \label{eq:tmp392} \\
    &= 1 - \frac{\sum_{y=1}^{\theta-1} y [\Ts^{y-1}\tv_0]_1 - (\theta-1) [\Ts^{\theta-1} \mathbf{1}_{E+1}]_1}{[(\Is_{E+1}-\Ts)^{-1} \mathbf{1}_{E+1}]_1},\label{eq:AVP}
\end{align}
where~\eqref{eq:tmp648} holds because within the $i$th  inter-refresh period, the AoI exceeds $\theta$ in the last $(y_i-\theta+1)^+$ slots (see Fig.~\ref{fig:AoI_process}); \eqref{eq:tmp649} holds because $\frac{1}{m}\sum_{i=1}^m y_i \to \E{Y}$ and $\frac{|\{i \in [m]\colon y_i = y\}|}{m} \to \P{Y = y}$ as $m\to \infty$.


\section{Numerical \hoang{Experiment}} \label{sec:results}

\subsection{Channel \cgd{Model} and Successful Delivery Probability} \label{sec:succ_prob}
We assume that a slot comprises $n$ uses of a real-valued \gls{AWGN} channel. In a slot, active device~$i$ with battery level $b_i$ transmits a signal $\cgd{\sqrt{\frac{b_i}{n}}}\xv_i \!\in\! \RR^n$ ($\|\xv_i\| \!=\! 1$) with power $b_i/n$. The received signal is 
$
    \yv = \sum_{i=1}^K \sqrt{b_i/n} \; \xv_i + \zv,
$
where $K$ is the number of active devices 
and $\zv \sim \Nc(0,\sigma^2)$ is the \gls{AWGN}. 
The devices transmit at rate $R$ bits/channel use, i.e., $\xv_i$ belongs to a codebook containing $2^{nR}$ codewords. We consider shell codes for which the codewords are uniformly distributed on the unit sphere. We \cgd{analyze} two decoding scenarios.

\subsubsection{Without Capture} 
In this scenario, all collided packets are lost. Decoding is attempted only on packets transmitted in singleton slots. This model allows us to revisit the collision channel model commonly used in modern random-access analyses, and further account for single-user decoding errors due to finite-blocklength effects. 
Consider an active device that transmits with $b$ energy units \cgd{and assume that} the battery profile of the remaining $U-1$ devices is $\Lm$. The successful delivery probability of the device of interest is
\begin{equation}
    w_{b,\Lm} = (1-\epsilon_b)\textstyle\prod_{i=0}^E(1-\pi_i)^{L_i}, \label{eq:w_noCapture}
\end{equation}
where $\epsilon_b$ is the error probability of decoding the device of interest in a singleton slot. To compute $\epsilon_b$, we use that 
the maximum achievable rate 
is~\cite[Th.~54]{Polyanskiy2010}
\begin{equation} \label{eq:rate_FBL}
    R^* = C(b) - \sqrt{\tfrac{V(b)}{n}} Q^{-1}(\epsilon_b) + O\big(\tfrac{\ln n}{n}\big)
\end{equation}
where $C(b) = \frac{1}{2}\log_2\big(1+\frac{b}{n\sigma^2}\big)$, $Q^{-1}(\cdot)$ is the inverse of the Gaussian Q-function $Q(z) = \frac{1}{2\pi}\int_{z}^\infty e^{-t^2/2} {\rm d} t$, and $V(b) = \frac{\frac{b^2}{n^2\sigma^4} + 2 \frac{b}{n\sigma^2}}{2(\frac{b}{n\sigma^2}+1)^2}\log_2^2(e)$ is the channel dispersion. 
For a fixed rate~$R$, we use~\eqref{eq:rate_FBL} to approximate $\epsilon_b$ as
$
    \epsilon_b \approx Q\Big(\sqrt{\frac{n}{V(b)}} (C(b) - R) \Big),
$
where we omitted $O(\frac{\ln n}{n})$, which is negligible for large $n$. 

\subsubsection{With Capture} In this case, the receiver attempts to decode every packet transmitted in a slot by treating all other colliding packets as noise.  Consider an active device with battery level~$b$ and let the battery profile of the remaining $U-1$ devices be~$\Lm$. Furthermore, assume that out of the other $L_i$ devices with battery level $i$, $\bar{L}_i$ devices transmit. Then the interference-to-noise power ratio is $\tilde{P} = \frac{1}{n\sigma^2}\sum_{i=0}^E i \bar{L}_i$, and the signal-to-interference-plus-noise ratio is $\bar{P} = \frac{b/(n\sigma^2)}{\tilde{P} + 1}$. In this setup, \cgd{an} achievable rate for the device of interest is given as in~\eqref{eq:rate_FBL} with $C(b)$ and $V(b)$ replaced by $\frac{1}{2}\log_2(1+\bar{P})$ and 
    $V'(b,\{\bar{L}_i\}) = \frac{\frac{b^2}{n^2\sigma^4}(1 \!+\! 2 \tilde{P} \!+\! \tilde{P}^2 \!-\! \breve{P}) + 2\frac{b}{n\sigma^2}(\tilde{P}\!+\!1)^3}{2(\tilde{P}+1)^2(b /(n\sigma^2) + \tilde{P} + 1)^2} \log_2^2 e$,
respectively~\cite[Th.~2]{Scarlett2017}. Here, $\breve{P} = \frac{1}{n^2\sigma^4}\sum_{i=0}^E i^2 \bar{L}_i$. Given $\{\bar{L}_i\}_{i=0}^E$, the error probability of the device is approximated as
\begin{equation}   
    \epsilon_{b,\{\bar{L}_i\}} \approx 
    Q\Big(\sqrt{\tfrac{n}{V'(b,\{\bar{L}_i\})}} \big(\tfrac{1}{2}\log_2(1\!+\!\bar{P}) \!-\! R\big) \Big).
\end{equation}

We further assume that the receiver employs \gls{SIC}. It first decodes all devices that transmit with $E$ energy units, removes the decoded packets, then decodes all devices that transmit with $E-1$ energy units, and so on. We assume that \cgd{the decoding of} a packet of energy $j$ \cgd{is attempted only} if all higher-energy packets have been correctly decoded and removed. In this case,
the battery profile of the interfering devices becomes $\hat{\Lm}\of{j} = (\hat{L}\of{j}_0, \dots, \hat{L}\of{j}_E)$, where $\hat{L}\of{b}_i = \bar{L}_i \ind{i \le b}$; 
if $j > b$, then $\hat{L}\of{j}_i$ takes value $0$ if $i > j$,  value $\bar{L}_i - 1$ if $i = j$, value $\bar{L}_i + 1$ if $i = b$, and value $\bar{L}_i$ if $i < j, i \le b$. 
It follows that 
\begin{equation}
    w_{b,\Lm} = \mathbb{E}_{\{\bar{L}_i\}_{i=1}^E}\Big[\big(1 - \epsilon_{b,\hat{\Lm}\of{b}}\big)\textstyle\prod_{j > b}\big(1 - \epsilon_{j,\hat{\Lm}\of{b}}\big)^{\bar{L}_j}\Big], \label{eq:w_capture}
\end{equation}
where $\bar{L}_i$ follows the binomial distribution with parameters $(L_i,\alpha\pi_i)$. Note that $\hat{\Lm}\of{b}$ is a function of $(b,\{\bar{L}_i\}_{i=1}^E)$.

Hereafter, we consider a slot length $n$ of $100$ channel uses and transmission rate $R$ of $0.8$ bits/channel use. 

\subsection{\hoang{\gls{AoI} and Throughput Evaluation}}

We first verify the accuracy of the exact and approximate AoI analysis by comparing with simulation results obtained from an implementation of the complete protocol operations over $10^7$ slots. 
To enable the computation of the exact average \gls{AoI}, we consider a small system with $U = 30$ and $E = 2$. We further set $\eta = 0.05$, $\theta = 1000$, and $\sigma^2 = -20$~dB. 
In Fig.~\ref{fig:sim_vs_analytic}, we plot the average AoI (normalized by $U$) and \gls{AVP} for the considered setting with capture. We consider 
$\piv=(1, 1)$ and $\piv = (0, 1)$. 
In both cases, the approximate average \gls{AoI}~\eqref{eq:avgAoI} matches well both the simulation results and exact analytical results. The approximate \gls{AVP}~\eqref{eq:AVP} is also in agreement with the simulation result. This confirms that our approximate analysis provides an accurate prediction of the \gls{AoI} performance.  

\begin{figure}[t!]
    \centering
    \input{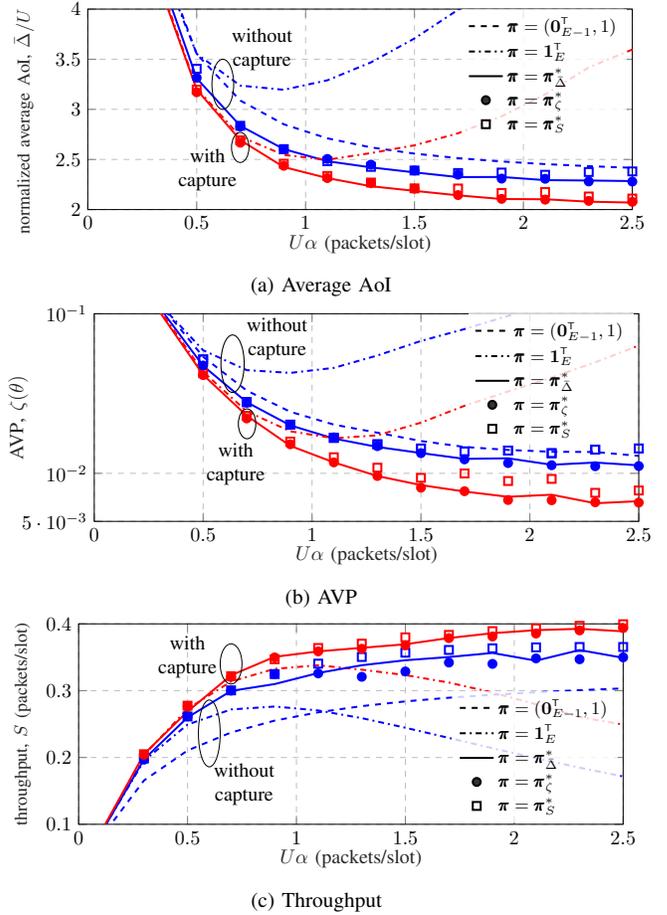}
    \caption{\hoang{Approximate} average AoI, \gls{AVP}, and throughput vs. the total number of new updates in a slot ($U\alpha$) for different values of the transmission probabilities $\piv$. \cgd{With $\piv = (\mathbf{0}^\T_{E-1}, 1)$, the performance with capture coincides with that without capture.} 
    	Here, $U = 1000$, $\eta = 0.005$, $E = 8$, $n = 100$, $R = 0.8$, $\theta = 10000$, and $\sigma^2 = -20$~dB.}
    \label{fig:opt_pi_eta.005}
    \vspace{-.3cm}
\end{figure}

Next, we report the approximate average \gls{AoI} and \gls{AVP} for a larger system with $U = 1000$, $E=8$, $\eta = 0.005$, and $\theta = 10000$.  We optimize $\piv$ to obtain 
	$\piv^*_{\bar{\Delta}} = \argmin_{\piv \in [0,1]^E} \bar{\Delta}$, 
	$\piv^*_{\zeta} = \argmin_{\piv \in [0,1]^E} \zeta(\theta)$, 
	$\piv^*_{S} = \argmax_{\piv \in [0,1]^E} S$.
We numerically solve these optimization problems using the Nelder-Mead simplex algorithm~\cite{nelder1965simplex}. 
In Fig.~\ref{fig:opt_pi_eta.005}, we plot the minimized average \gls{AoI}, minimized \gls{AVP}, and maximized throughput as a function of $U\alpha$, and compare it with two baseline strategies:
i) $\piv = (\mathbf{0}^\T_{E-1}, 1)$, i.e., a device  transmits only with full battery, and ii) $\piv = \mathbf{1}_E^\T$, i.e., a device transmits a new update whenever possible.\footnote{\hoang{The strategy with $\piv = \mathbf{1}_E^\T$ \cgd{(best-effort uniform policy)} was shown to be average-AoI optimal in the case of infinite battery capacity for the single-source scenario with perfect or packet-erasure channel~\cite{Wu2018optimal,Feng2021}.}} We see that the optimized $\piv$ leads to significant improvement in all three metrics. The strategy $\piv = \mathbf{1}_E^\T$ is close to optimal when $U\alpha$ is small, especially with capture. However, this strategy becomes highly suboptimal when $U\alpha$ increases, since it causes many collisions. In contrast, the strategy with $\piv = (\mathbf{0}^\T_{E-1}, 1)$ has a decreasing gap to the optimal performance without capture when $U\alpha$ is large. However, this strategy does not benefit from decoding with capture since the level of interference in non-singleton slots is always high. 
With capture, the minimized average AoI and maximized throughput are improved by about $10\%$ and $18.7\%$, respectively, for $U\alpha = 2.5$, compared to decoding without capture.

The optimized $\piv$ can be different for different metrics. \hoang{While the $\piv$ optimized for the average AoI also performs \hoang{close to optimal} for the \gls{AVP} and vice versa, \cgd{the one optimized for throughput} leads to suboptimal average \gls{AoI} and \gls{AVP}, especially for high $U\alpha$.} 
Without capture, \cgd{for the optimized $\piv$, devices with higher battery level transmit more often}. 
With capture, the devices transmit with either low or high power, facilitating the decoding of high-energy packets and then \cgd{of} low-energy packets after \gls{SIC}. For example, for $U\alpha = 2.1$, $\piv^*_{\bar{\Delta}}$ is $(0, 0, 0, 0.68, 1, 1, 1, 1)$ without capture and $(0, 0, 1, 1, 0, 0, 0, 1)$ with capture.

\vspace{-.1cm}
\section{Conclusions} \label{sec:conclusions}
We studied the impact of energy harvesting on information freshness in slotted ALOHA networks. Leaning on a Markovian analysis, we provided an exact analytical analysis of the average AoI, as well as an approximate analysis that results in easy-to-compute and accurate approximations of both the average AoI and \gls{AVP}. 
Our main findings are as follows: i) transmitting a new update whenever possible is beneficial only for low update generation rates, while waiting for sufficient energy before transmitting is preferable for high update generation rates, ii) significant gains with respect to these \cgd{two} baseline strategies can be achieved with an optimized strategy, iii) \hoang{the \gls{AVP}-minimizing strategy performs close to optimal in terms of the average \gls{AoI} and vice versa, while \cgd{the one optimized for} throughput entails a notable loss in terms of the \gls{AoI} metrics,} iv) decoding with capture significantly outperforms decoding without capture, especially for high update generation rates. 

\bibliographystyle{IEEEtran}
\bibliography{IEEEabrv,./biblio}

\end{document}